# Skyrmion nucleation via localized spin current injection in confined nanowire geometry in low chirality magnetic materials

Sourav Dutta [*1], Dmitri E. Nikonov [2], George Bourianoff [3], Sasikanth Manipatruni [2], Ian A. Young [2], and Azad Naeemi [1]

[1]School of Electrical and Computer Engineering, Georgia Institute of Technology, Atlanta, GA 30332 USA

[2]Components Research, Intel Corporation, Hillsboro, OR 97124 USA

[3]Intel Corporation, 1300 S. MoPac Exp, Austin, TX 78746 USA

*Correspondence to sdutta38@gatech.edu

**Magnetic skyrmions have been the focus of intense research with promising applications in memory, logic and interconnect technology. Several schemes have been recently proposed and demonstrated to nucleate skyrmions. However, they either result in an uncontrolled skyrmion bubble production or are mostly targeted towards integration with racetrack memory device. In this work, we propose a novel scheme for a controlled single skyrmion nucleation in a confined nanowire geometry with sub-100 nm width using a generalized approach of "localized spin current injection" technique in material systems exhibiting low Dzyaloshinskii–Moriya interaction (DMI). Our proposed nucleation mechanism follows a pathway involving the creation of a reversed magnetic domain containing one or more pairs of vertical Bloch lines (VBLs) that form an edge-to-edge domain wall as the VBLs get annihilated at the edge of the nanowire. However, pinning of the edge domain walls within a narrow gap using notches or anti-notches results in the creation of a magnetic bubble with defect-free domain wall that eventually relaxes into a circular skyrmion structure. Our simulations predict that the proposed mechanism allows skyrmion nucleation on sub-nanosecond timescale, shows robustness to variations like local pinning sites and is applicable for any skyrmion-based logic, memory and interconnect application.**

Recently, information processing utilizing magnetic quasi-particles such as magnons [1-5], domain walls [6, 7] and skyrmions [8-11] has been the focus of intense research for low-power, non-volatile beyond-CMOS logic, memory and interconnect application. Magnetic skyrmions are particle-like magnetic configurations originating from a chiral interaction known as the Dzyaloshinskii–Moriya interaction (DMI) [12, 13], characterized by a topological skyrmion number or winding number (equal to -1) defined as $S = \frac{1}{4\pi} \int \vec{m} \cdot \left( \frac{\partial \vec{m}}{\partial x} \times \frac{\partial \vec{m}}{\partial y} \right) dxdy$, where $\vec{m}$ is the normalized magnetization vector. Skyrmions have been observed in non-centrosymmetric magnetic materials like MnSi [9, 10, 14], FeCoSi [15, 16], FeGe [17, 18] and MnFeGe [19] and in multilayers of heavy metal exhibiting high spin-orbit coupling and ultrathin magnetic film with broken inversion symmetry like Ir/Fe [11, 20], Pt/CoFe [21], Ta/CoFe [21],Pt/Co/Ta [22],Pt/Co/Ir [23], Pt/CoFeB/MgO [24, 25], Ta/CoFeB/TaO$_x$ [26, 27] and Ta/CoFeB/MgO [28]. Contrary to the ferromagnetic exchange interaction that favors collinear arrangement of spins, DMI allows the stabilization of helical magnetic order [8-11, 29, 30]. The advantages of utilizing magnetic skyrmions for information processing stems from their topological stability [31], smaller size [20, 32, 33] compared to magnetic bubbles stabilized by magnetostatic interaction, and motion induced by current densities as low as $10^6$ A/m$^2$ see refs. [14, 32, 34-37]. In terms of application, the main focus of research so far has been on the development of a dense, low-power skyrmion-based racetrack memory [24, 32, 37-39], similar to the domain-wall racetrack memory [6] proposed earlier by IBM. In a skyrmion-based racetrack memory, the stream of binary data (sequence of "1"s and "0"s) can be represented as the presence or absence of a skyrmion, respectively. A new proposal for encoding the binary data using two different particle-like states: skyrmions and chiral bobbers [40, 41] have also been recently demonstrated [42]. Besides showing promise in memory technology, recent works have also highlighted the development of skyrmion-based nano-oscillator [43], logic gate [44], transistor [45] and reservoir computing [46]. However, a key application of skyrmions for interconnecting magnetic logic remains unexplored. Relying on the small size, high-packing density, displacement via ultra-low current density, and robustness to defects and pinning, a skyrmion-based magnetic interconnect can provide a low-power, high throughput and robust alternative to other candidates for interconnecting magnetic logic [47, 48] without using an electrical transduction for charge to spin signal conversion.

A key ingredient for any skyrmion-based memory, logic or interconnect is an efficient, reliable and controlled scheme of nucleation or writing of magnetic skyrmions with reduced complexity. Several techniques have since been proposed for skyrmion nucleation [36, 38, 49-54]. However, they either require low temperature [49], provide an uncontrollable random production of skyrmion bubbles [50, 55, 56] at available DMI, require a considerably large DMI for robust transformation of domain wall to skyrmions [54, 57] or require engineered interfacial perpendicular magnetic anisotropy [28]. An earlier proposal by Iwasaki et al. [36] and recent demonstrations by Buttner et al. [52] and Everschor et al. [51] have provided alternative controlled and integrated methods of skyrmion nucleation using an electric current. However, they utilize the same stimulus (electric current) for skyrmion nucleation and dynamics, relying on the magnitude of the applied current pulse to either write or move skyrmions, and as such are mostly suited for a skyrmion-based racetrack memory technology.

In this work, we revisit and explore the nucleation mechanism proposed by Sampaio et al. [32] using localized spin-polarized current injection. A "localized spin current injection" based nucleation technique allows decoupling the writing from the driving mechanism and provides a generalized and universal scheme for skyrmion creation that can be used both for a racetrack memory, logic application or interconnect. Fig. 1 shows an illustration of a possible implementation of skyrmion interconnect used for connecting magnetic logic (which can be a traditional all-spin-logic (ASL) [58], modified magnetostriction-assisted ASL [59] or spin wave logic device [5]). As shown in Fig. 1, the last stage of the magnetic logic can be connected to the skyrmion interconnect using a conventional spin valve structure. Depending on the orientation of the ferromagnetic layer (last stage of the magnetic logic), spin current with polarization $\sigma_{spin}$ = +z or –z can be injected into the magnetic layer of the skyrmion interconnect. A +z polarized spin current retains the uniform ferromagnetic state of the nanowire while a –z polarized spin current aids in the nucleation of a magnetic skyrmion. Hence, the stream of binary data (sequence of "1"s and "0"s) can be represented as the presence or absence of a skyrmion. The focus of this work is on the creation of stable skyrmion in such a nanowire geometry in the presence of low chirality. The spin-orbit torque driven skyrmion motion and performance evaluation of skyrmion interconnect will be the subject of a separate discussion and lies beyond the scope of this work.

The pathway towards a successful skyrmion nucleation using the "localized spin current injection" technique involves: (a) creation of a magnetic domain with reversed magnetization surrounded by a pseudo-domain wall containing one or more pairs of vertical Bloch lines (VBLs), (b) eventual annihilation of the VBLs creating a magnetic bubble with defect-free domain wall and (c) relaxation of the magnetic bubble into a circular skyrmion structure. According to the simulations by Sampaio and colleagues [32], such a skyrmion nucleation using spin injection requires a high DMI value (D ≥ 5 mJ/m² in their simulation). Other simulation works involving such nucleation technique have also considered a material system with a high DMI value of D = 3 mJ/m² [60-62]. While a higher value of DMI is sought for, recent experimental investigations have suggested a maximum obtainable value of 2 mJ/m² and 4 mJ/m² for Ir/Co/Pt [63] and Ir/Fe [11] systems, respectively. In more beyond-CMOS technologically relevant spintronic materials like Pt/CoFeB [24, 25, 64] and Ta/CoFeB [27, 38, 50], the obtained DMI is 1-1.5 and 0.1-0.5 mJ/m², respectively. In this work, we propose and investigate a controlled single skyrmion nucleation mechanism in a confined nanowire geometry with sub-100 nm width with a material system (Pt/CoFeB or Ta/CoFeB) displaying a lower DMI (< 2 mJ/m²). The proposed mechanism hinges on the pinning of the edge-to-edge domain wall using notches or anti-notches as the VBLs get annihilated at the edge of the nanowire that facilitates the re-generation of a skyrmion bubble with defect-free domain wall.

### Skyrmion nucleation in nanowire via spin current injection

We start by revisiting the process of skyrmion nucleation proposed by Sampaio et al. [32] in an ultrathin narrow nanowire geometry as shown in Fig. 2(a). We consider a multilayer stack of ferromagnetic layer sandwiched between a heavy metal exhibiting high spin-orbit coupling and an oxide layer. The interface between the heavy metal and the ferromagnet generates an interfacial DMI of the form $E_{DMI} = \sum_{i,j} d(u_{i,j} \times \hat{z}) \cdot (\vec{S_i} \times \vec{S_j})$ where $\vec{S_i}$ is the atomic spin unit vector, d is the DMI coupling energy, $u_{i,j}$ is the unit vector connecting sites $i$ and $j$, and $\hat{z}$ is the unit vector along the out-of-plane direction as shown in Fig. 2(a). Additionally, such a multilayer also displays a strong perpendicular magnetic anisotropy (PMA). We choose the ferromagnetic material parameters corresponding to ultrathin CoFeB while the DMI strength is varied between 0.5 – 3 mJ/m² (that corresponds to a choice of Pt, Ta and Ir as the heavy metal). Detailed simulation procedure is provided in the methods section. Since the stabilization of an isolated skyrmion depends on the interplay between the dipolar, exchange, DMI and anisotropy energy, we correspondingly vary the perpendicular anisotropy $K$ between 0.26 – 0.9 MJ/m³ as shown in Fig. 2(b). For a given $K$, the skyrmion size increases with increasing DMI (till the critical value $D_C = \frac{4}{\pi}\sqrt{AK_{eff}}$), where $K_{eff}$ is the effective out of plane anisotropy, while for a fixed DMI, increasing $K$ reduces the skyrmion diameter. For the rest of our simulations, we chose the combination of $D$ and $K$ values to be (0.5 mJ/m$^2$, 0.26 MJ/m$^3$), (1.0 mJ/m$^2$, 0.35 MJ/m$^3$), (1.5 mJ/m$^2$, 0.45 MJ/m$^3$), (2.0 mJ/m$^2$, 0.6 MJ/m$^3$), (2.5 mJ/m$^2$, 0.8 MJ/m$^3$) and (3.0 mJ/m$^2$, 0.9 MJ/m$^3$) that correspond to a skyrmion diameter of 20 – 30 nm.

We consider a spin current injection through a circular nano-contact as shown in Fig. 2(a) for skyrmion nucleation. The nanowire is initially magnetized ($m_{FM}$) in the +z direction. The spin current with spin polarization ($\sigma_{spin}$) in the –z direction is injected through the circular nano-contact region of diameter 20 nm for a time period of 1 ns. Throughout this work, the field-like torque is assumed to be 0; however, including the field-like torque in simulations does not change the main conclusion of the work. At low spin current density ($J_S$), no domain reversal is observed and the initial ferromagnetic state is retained. Above a critical threshold ($J_S \geq 1.5 \times 10^{12}$ A/m$^2$), we see the formation of a reversed magnetic domain. However, the possibility for a successful skyrmion nucleation depends highly on the strength of the DMI [32] as depicted in Fig. 2(c). To better elucidate the reason behind the two distinct regions of success and failure, we investigate in details the skyrmion nucleation mechanism.

As shown in Fig. 2(d) for the case of large DMI (3 mJ/m$^2$) and injected spin current density $J_S = 2 \times 10^{12}$ A/m$^2$, at t = 0.26 ns, the magnetization inside the nano-contact region is reversed with magnetization pointing in the –z direction. This magnetic configuration corresponds to a magnetic bubble surrounded by a pseudo-domain wall. The pseudo-domain wall is composed of two regions with opposite chirality (left-handed with spins pointing outward and right-handed with spins pointing inward) separated via a pair of vertical Bloch lines (VBLs). The corresponding distribution of DMI energy density is shown in Fig. 2(e). As the spins inside the pseudo-domain wall reorient rapidly, the region with chirality favored by DMI (left-handed in our case) expands while the region with opposite chirality shrinks into a small defect-like region trapping a large positive DMI energy density as shown at t = 0.317 and 318 ns in Fig. 2(d, e, f). As the size of this defect-like region similar to a Bloch point rapidly decreases, eventually the large positive DMI energy density forces the VBLs to get annihilated with rapid reorientation of the spins and emission of radial spin waves as seen at t = 0.319 ns in Fig. 2(d, e). The emission of the spin waves carries off energy that ultimately explains the stability of the skyrmion through the creation of the potential well necessary for skyrmion stability and will be the subject of a separate work. Fig. 2(g) shows the change in the total energy of the system, overcoming the topological energy barrier. As the VBLs get annihilated, the skyrmion number drastically changes from 0 to a value close to -1 corresponding to an isolated magnetic skyrmion as shown in Fig. 2(g).

The case of a lower DMI (1 mJ/m$^2$) presents a different scenario. As shown in Fig. 2(h), upon spin current injection, VBLs are created. However, as the spins inside the pseudo-domain walls reorient in the direction favored by the DMI, the VBLs gets attracted towards the edge of the nanowire due to the attractive interaction with the inward tilted edge magnetization as seen at t = 0.23 in Fig. 2(h). Eventually, the VBLs are expelled and the magnetic bubble transforms into an edge-to-edge domain wall that slowly expands to form a multi-domain state as seen at t = 0.3 and 0.4 ns. As the current pulse stops after 1 ns, the domain wall collapses to form the initial uniform ferromagnetic state. Fig. 2(i) shows the change in the total energy density and the skyrmion number.

**Skyrmion nucleation in nanowire with edge material**

To circumvent the problem of VBLs getting attracted towards the edge giving rise to domain wall formation, next we incorporate an additional potential barrier at the two edges of the nanowire as shown in Fig. 3(a) using a 5 nm wide edge material with high PMA. As an illustrative example, we consider the material parameters for the edge material equal to that of Samarium Cobalt SmCo$_5$ (Ms = 0.84 MA/m$^3$, A = 12 pJ/m, K = 5 – 17 MJ/m$^3$, D = 0.1 mJ/m$^2$) [65]. Similar to the previous case, the nanowire initially has a uniform magnetization in the +z direction while the spin current with spins polarized in the –z direction is injected through the circular nano-contact region of diameter 30 nm for 1 ns long. We mainly focus on the range of DMI that previously resulted in an unsuccessful skyrmion nucleation owing to the formation of edge-to-edge domain wall. As shown in Fig. 3(b), we can now successfully nucleate skyrmions with lower DMI (< 3 mJ/m$^2$) owing to the presence of the edge material. However, the required spin current density remains high. Fig. 3(c) shows the skyrmion nucleation mechanism with injected spin current density $J_S = 10 \times 10^{12}$ A/m$^2$ in the presence of a DMI of 2 mJ/m$^2$. The pathway for successful skyrmion nucleation remains the same as before. At t = 0.12 ns, the magnetization inside the nano-contact region is reversed forming a magnetic bubble surrounded by a pseudo-domain wall containing a pair of VBLs. The repulsive force from the edges with high PMA prevents the VBL from getting expelled at the edge of the nanowire. The spins inside the pseudo-domain wall rapidly reorient at t = 0.15 ns as seen in Fig. 3(c) expanding the region with chirality favored by DMI while shrinking the region with opposite chirality into the defect-like regions trapping a large positive DMI and exchange energy density as seen in Fig. 3(d). At t = 0.16 ns, the size of this defect-like region decreases to a minimum trapping a large amount of positive DMI energy density that forces the VBLs to get annihilated with rapid reorientation of the spins. As the VBL gets annihilated at t = 0.164 ns, the bubble relaxes to a skyrmion configuration with skyrmion number changing to a value close to -1 as seen in Fig. 3(e) with a drastic reduction in the energy of the system.

However, for even lower DMI D = 0.5 mJ/m$^2$, we do not see an annihilation of the VBLs. On the contrary, the VBL pair exists throughout the duration of the applied spin current pulse as see in Fig. 3(f). Additionally, we see that as the position of the VBLs change due the applied spin torque, the magnetic bubble relaxes, elongating in the longitudinal direction, as seen at t = 0.3 ns when the defect-like point aligns along the length of the nanowire. This also results in a decrease in the maximum positive DMI energy density concentrated at the defect-like regions of the pseudo-domain wall as seen in Fig 3(g). The positive DMI energy density attains its peak value when the defect-like point aligns along the transverse (y) direction as seen at t = 0.6 ns in Fig. 3(f, g). However, this peak value of maximum positive DMI energy density remains unsuccessful in annihilating the VBLs. This points to the fact that there exists a threshold DMI strength or corresponding trapped positive DMI energy density that can force the spins inside the defect-like regions to reorient along the direction favored by DMI, thus annihilating the VBLs. As the current pulse stops after 1 ns, the magnetic bubble collapses to form the initial uniform ferromagnetic state. Fig. 3(h) shows the change in the total energy and the skyrmion number.

We additionally study the impact of local variations in magnetic parameters which may lead to pinning and annihilation of VBLs by using a local site-by-site variation of both PMA and DMI. The detailed simulation process incorporating variation and the result are highlighted in the supplementary information section S1. We do not see any significant variation in the result from that highlighted in Fig. 3(b).

## Skyrmion nucleation in nanowire with gap in edge material

Next, we investigate a novel mechanism of skyrmion nucleation via VBL expulsion while pinning the edge-to-edge domain wall using a narrow gap in the edge material. The proposed geometry is illustrated in Fig. 4(a). The optimum choice of the gap length for successful skyrmion nucleation, as shown in Fig. 4(b), is explained later. Fig. 4(c) shows the skyrmion nucleation process with injected spin current density $J_S = 5 \times 10^{12}$ A/m$^2$ in the presence of D = 1mJ/m$^2$ and a gap length of 20 nm. At t = 0.22 ns, the magnetization inside the nano-contact region is reversed forming the magnetic bubble containing the VBL pair. The position of the VBLs changes due the applied spin torque and as it aligns with the edge gap at t = 0.93 ns, the VBL gets attracted towards the edge initiating the formation of edge-to-edge domain wall. However, due to the presence of edge materials with high PMA on both sides of the gap, the edge domain wall gets pinned contrary to the case without edges as seen in Fig. 2(h). The impact of the edge material with high PMA is similar to that of trapping a domain wall pair via pinning using notches or anti-notches [66, 67]. As the edge domain wall remains pinned forming a fractional edge skyrmion at t = 0.95 ns, the spins at the center of the gap where the net magnetization points in the –z direction (blue region in Fig 4(c)) tilt outward due the boundary condition imposed by DMI $\frac{d\vec{m}}{dn} = \frac{D}{2A}(\hat{z} \times \hat{n}) \times \vec{m}$ [68], where $\hat{n}$ is the edge normal. This rapid reorientation of the spins in the gap regions along with the simultaneous pinning of the edge domain wall, prohibiting domain wall propagation, results in the re-generation of a magnetic bubble surrounded by a pseudo-domain wall but without any VBL (all spins pointing outward) as seen in zoomed-in picture of the gap region in Fig. 4(d). Eventually, the bubble decouples from the edge-gap and moves towards the center of the nanowire due the repulsive force from the edges and relaxes into a stable skyrmion configuration accompanied by a change in the skyrmion number to a value close to -1 and reduction in the total energy as shown in Fig. 4(e). As the spin current pulse stops after 1 ns, the skyrmion size shrinks to its intrinsic diameter and remains stable as seen at t = 2 ns. As highlighted in Fig. 4(e), this mechanism allows sub-nanosecond skyrmion creation.

We further analyze the impact of the gap length in order to obtain an optimum range for successful skyrmion nucleation. We first consider the case of an extremely narrow gap length of 5 nm as shown in Fig. 4(f). At t = 0.22 ns, the magnetic bubble forms and the position of the VBLs changes. However, as the VBL pair reaches the narrow gap at t = 0.9 - 0.92 ns, it moves past it without forming fractional edge skyrmion. This points to the fact that the minimum gap length must be at least equal to the domain wall width $\Delta_{DW} = \sqrt{A/K_{eff}}$ to support the two half domain walls (with the width of $\Delta_{DW}/2$ each) initiating the VBL annihilation process as shown in Fig. 4(g). Fig. 4(h) shows the change in the total energy and the skyrmion number remaining at 0. For the other extreme case of a wide gap, as shown in Fig. 4(i) for a gap length of 40 nm, the VBL gets attracted towards the edge gap. On reaching the edge, the bubble transforms into an edge-meron or half skyrmion (with topological change Q = - ½ as seen in Fig. 4(j)) as seen at t = 0.26 ns. The edge-meron either remains as an energetically meta-stable state or transforms into a pair of domain walls trapped between the anti-notches as seen at t = 1 ns. When the current pulse is turned off after 1 ns, the spin configuration collapses forming a uniformly magnetized nanowire as seen at t = 2 ns. Fig. 4(j) shows the change in the total energy and the skyrmion number. The scenario when the edge-meron or domain wall pair becomes energetically more favorable than the skyrmion occurs when the gap length becomes greater than the size of the skyrmion, putting a limitation on the optimum gap length. Fig. 4(b) shows the optimum gap length for successful skyrmion nucleation for different values of DMI and K. The upper and lower limits of optimum gap length obtained from micromagnetic simulations for all (D, K) pairs agree well with the above stated hypothesis. Similar results (not shown here) were obtained on reducing the spin current density to the threshold value of $J_S \sim 1.5 \times 10^{12}$ A/m$^2$.

**Robustness to variability**

Recent experiments on the existence of room-temperature skyrmions in sputtered multilayers [23, 24, 50, 63] have observed the presence of local variations of magnetic parameters which leads to pinning and disordered motion of skyrmions. Since in ultrathin films, local variations of thickness can dramatically change the thickness-dependent PMA and DMI, to incorporate such inhomogeneity, we use a local site-by-site uncorrelated variation of both PMA and DMI. Figs. 5(a) and (b) show an example of spatial variation of PMA and DMI used in the simulation, respectively. We use a normal distribution around a mean value of $K_0$ and $D_0$ with a standard deviation ($\sigma_K$ and $\sigma_D$) of 10% as shown in Figs. 5(a) and (b). The probability distribution plot of nucleating a skyrmion for different values of DMI (and PMA) and gap length, highlighted in Fig. 5(c), do not show any significant variation from Fig. 3(b), thus illustrating the robustness of the nucleation process to 10% variability in PMA and DMI. To further elucidate the reliability of the nucleation mechanism, we calculate the probability of nucleation for simultaneous variation in PMA and DMI ($\sigma_K$ and $\sigma_D$) up to 30% around the mean value of $K_0$ = 0.35 MJ/m$^3$ and $D_0$ = 1 mJ/m$^2$. As shown in Fig. 5(d), the probability falls below 0.9 beyond a large variation ($\sigma_K$ and $\sigma_D$) of 10%. Also, note that we consider a pessimistic case of site-by-site variation in material parameter over a mesh size (comparable to grain size in crystalline film like Co) of 1 nm with no correlation in variation between the neighboring cells. The impact of variation of other parameters like exchange interaction and saturation magnetization has been discussed in the supplementary information section S2.

**Skyrmion nucleation in nanowire with notches as pinning sites**

Having established that the edge material acts as pinning sites for edge domain wall and inspired by domain wall pinning at anti-notches, we next explore the possibility of skyrmion nucleation in a nanowire geometry in the presence of notches. The notches with dimensions 50 nm long and 5 nm wide are considered to be of the same material as the edges considered earlier (high PMA) that can act as pinning sites for the domain wall. Fig. 6(a) illustrates the proposed structure consisting of 4 rectangular notches (pinning site). The optimum gap length for successful skyrmion nucleation for different values of DMI and K using notches is shown in Fig. 6(b). Fig. 6(c) shows the skyrmion nucleation process with injected spin current density $J_S = 5 \times 10^{12}$ A/m$^2$ in the presence of D = 1mJ/m$^2$ and a gap length of 20 nm between the notches. The nucleation mechanism follows the same pathway as explained earlier with the formation of the magnetic bubble containing VBL as seen at t = 0.22 ns. At t = 0.63 ns, the VBL reaches the edge of the nanowire initiating the formation of the edge-domain wall. However, due to the presence of the notches at the edge of the nanowire, the domain wall gets pinned similar to Fig. 4(c, d) forming a fractional edge skyrmion while the spins at the center of the gap reorient to point in the outward direction as dictated by the boundary condition imposed by DMI. At t = 0.65 ns, a magnetic bubble re-generates surrounded by a pseudo-domain wall without any VBL. The bubble finally relaxes to a stable skyrmion structure. Fig. 6(d) shows the zoomed-in picture of the gap regions showing the formation of the defect-free magnetic bubble. Fig. 6(e) shows the change in the total energy and the skyrmion number highlighting sub-nanosecond skyrmion nucleation.

We further study the impact of the gap length between the notches or pinning sites on the skyrmion nucleation process. As shown in Fig. 6(f), a narrow gap of 10 nm does not allow the formation of the edge domain wall, prohibiting the expulsion of the VBLs at the edge of the nanowire. Hence the VBL pair moves past the gap without skyrmion nucleation. It is intriguing to find that the lower limit of gap length for successful skyrmion nucleation has increased to almost twice that of the previous case (see Fig. 6(b)). As shown in Fig. 6(g), the gap region between the notches or pinning sites now has to accommodate two domain wall widths $\Delta_{DW}$ for the VBL to reach the nanowire edge and get annihilated. Hence, the minimum gap length in this scenario increases to almost $2\Delta_{DW}$. Fig. 6(h) shows the change in the total energy and the skyrmion number remaining at 0. For a wider gap, as shown in Figs. 6(i), the bubble transforms into an edge-meron that becomes energetically more favorable than the skyrmion. The edge-meron exists as long as the current pulse remains on after which it gets annihilated at the edge of the nanowire. Fig. 6(k) shows the change in the total energy and the skyrmion number changing to a value of -1/2 denoting the creation of an edge-meron for the duration of the applied current pulse. We also see an increase in the upper limit of the optimum gap length (see Fig. 6(b)). This is due to an additional repulsive force applied by the notches or the pinning sites in the gap region on the two domain walls. As shown in Fig. 6(j) for a gap of 38 nm, this additional repulsion forces the two domain walls to move towards each other resulting in the re-generation of the defect-free magnetic bubble as seen at t = 1.02 ns in Fig. 6(j). The upper and lower limits of optimum gap length obtained from micromagnetic simulations for all (D, K) pairs as shown in Fig. 6(b) agree well with the above stated hypothesis. Similar results (not shown here) are obtained on reducing the spin current density to the threshold value of $J_S \sim 1.5 \times 10^{12}$ A/m$^2$.

## Conclusion

In conclusion, we have proposed and investigated a novel scheme of sub-nanosecond skyrmion nucleation in a confined nanowire geometry with sub-100 nm width using the generalized approach of "spin current injection" technique. Such a scheme is essential for skyrmion-based interconnect, and can also be applicable for skyrmion logic and memory applications. We perform micromagnetic simulations using experimentally demonstrated material parameters to demonstrate that the proposed scheme is technologically relevant for beyond-CMOS application (Pt or Ta with CoFeB). We consider material systems exhibiting a low value of DMI (< 2 mJ/m²). Our proposed mechanism follows a pathway involving (a) creation of a magnetic domain with reversed magnetization surrounded by a pseudo-domain wall containing one or more pairs of vertical Bloch lines (VBLs), (b) formation of edge-to-edge domain wall as the VBL gets attracted towards the edge of the nanowire, (c) pinning of the edge-to-edge domain wall using notches or anti-notches as the VBLs get annihilated at the edge of the nanowire and (d) creation of magnetic bubble with defect-free domain wall eventually relaxing into a circular skyrmion structure. Using the proposed mechanism, we were able to obtain successful skyrmion nucleation with spin current density as low as $J_S \sim 1.5 \times 10^{12}$ A/m². Note that for building a skyrmion interconnect, as shown in Figure 1, we consider this spin current to be injected via a conventional spin valve through a nano-contact of diameter around 20 nm. Hence, we estimate the net charge current to be in the range of 0.3 – 0.5 mA and the energy dissipation for skyrmion nucleation to be in the range of 2 – 5 fJ. We are aware that patterning magnetic materials at the nanometer scale discussed here will be challenging from a fabrication point of view, and may require careful localized techniques like using local ion beam and so on. But we believe that does not detract from the value of the theoretical understandings presented in this paper.

## Methods
### Micromagnetic simulation

We perform micromagnetic simulations in OOMMF [69] that numerically solves the Landau-Lifshitz-Gilbert (LLG) equation augmented with the spin-transfer-torque and interfacial DMI term [68] at zero-temperature. We consider the nanowire dimensions to be 200 nm long, 60 nm wide and 1 nm thick while the discretized cell size is taken to be 1 x 1 x 1 nm³. We choose the material parameters corresponding to ultrathin (1 nm) CoFeB with saturation magnetization $M_s$ = 0.65 MA/m [27, 50] and exchange stiffness A = 10 pJ/m [25, 27]. To explore the impact of DMI strength on the nucleation process, we vary DMI ($D$) between 0.5 – 3 mJ/m² (that corresponds to a choice of Pt, Ta and Ir as the heavy metal) and correspondingly also vary the perpendicular anisotropy ($K$) between 0.26 – 0.9 MJ/m³ to stabilize skyrmion.

To study the robustness of our proposed nucleation mechanism to variability, we perform numerical simulations using a local site-by-site uncorrelated spatial variation of material parameters like PMA, DMI, exchange stiffness and saturation magnetization. We use a normal distribution to model the variability with the standard deviation varying between 5-30% as has been highlighted in the main text and supplementary information section S1 and S2.

We perform addition simulations (see supplementary information section S3) for skyrmion nucleation in nanowire with gap in edge material using a higher value of saturation magnetization ($M_s$ = 1 MA/m) similar to that reported by other groups for thin film CoFeB [64, 70, 71]. We use the same range of DMI ($D$) as used in the main text between 0.5 – 2 mJ/m² (that corresponds to a choice of Pt and Ta as the heavy metal). Due to the increased value of Ms, the correspondingly perpendicular anisotropy ($K$) is chosen between 0.62 – 0.95 MJ/m³ to stabilize skyrmion and achieve a skyrmion diameter of 20 – 30 nm. The same qualitative result is obtained, similar to Fig. 3(b) of the main text.

**Author Contributions**

**Figures**

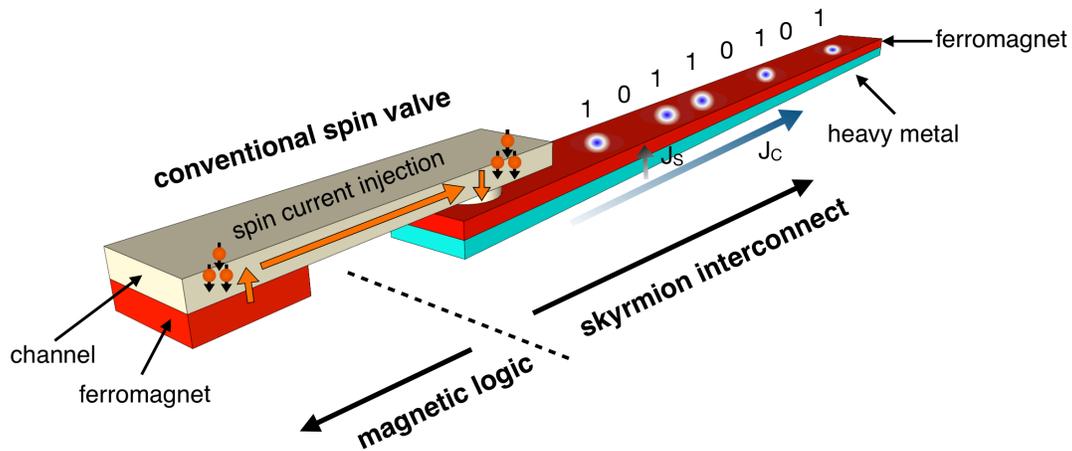

Fig. 1. Illustration of a skyrmion interconnect used for connecting a magnetic logic. The last stage of the magnetic logic can be connected to the skyrmion interconnect using a conventional spin valve structure. The orientation of the ferromagnetic layer (last stage of the magnetic logic) dictates the polarization of the spin current ($\sigma_{spin}$ = +z or –z) that is injected into the magnetic layer of the skyrmion interconnect. A +z polarized spin current retains the uniform ferromagnetic state of the nanowire while a –z polarized spin current nucleates a magnetic skyrmion. The stream of binary data (sequence of "1"s and "0"s) is represented as the presence or absence of a skyrmion in the interconnect.

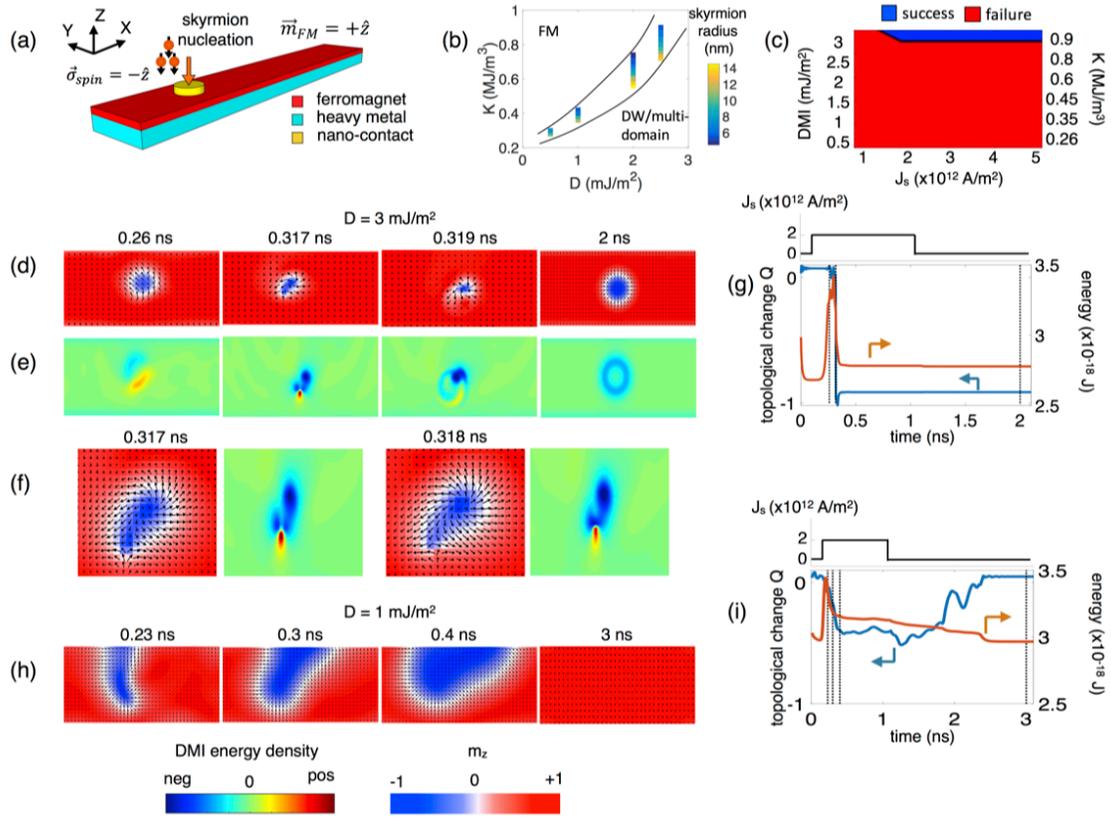

Figure 2. (a) Scheme of skyrmion nucleation using localized spin current injection into an ultrathin ferromagnetic nanowire sandwiched between a heavy metal and an oxide. The spin-polarized current is injected through a circular nano-contact. (b) Skyrmion stability for different values of DMI (D) and PMA (K). The color map shows the radius of the stabilized skyrmion. (c) Probability of skyrmion nucleation for different values of (D, K) and injected spin current density $J_s$. (d, e) Temporal evolution of magnetization and corresponding DMI energy density after spin current injection for D = 3 mJ/m$^2$. (f) Zoom-in of the spin current injection region showing a large positive DMI energy density concentrated within a small defect-like region that leads to the annihilation of VBLs and generation of skyrmion. (g) Change in the skyrmion number from 0 to -1 upon VBL annihilation and the change in the total energy overcoming the topological barrier. (h) Temporal evolution of magnetization after spin current injection for D = 1 mJ/m$^2$. (i) Corresponding change in the skyrmion number and the total energy.

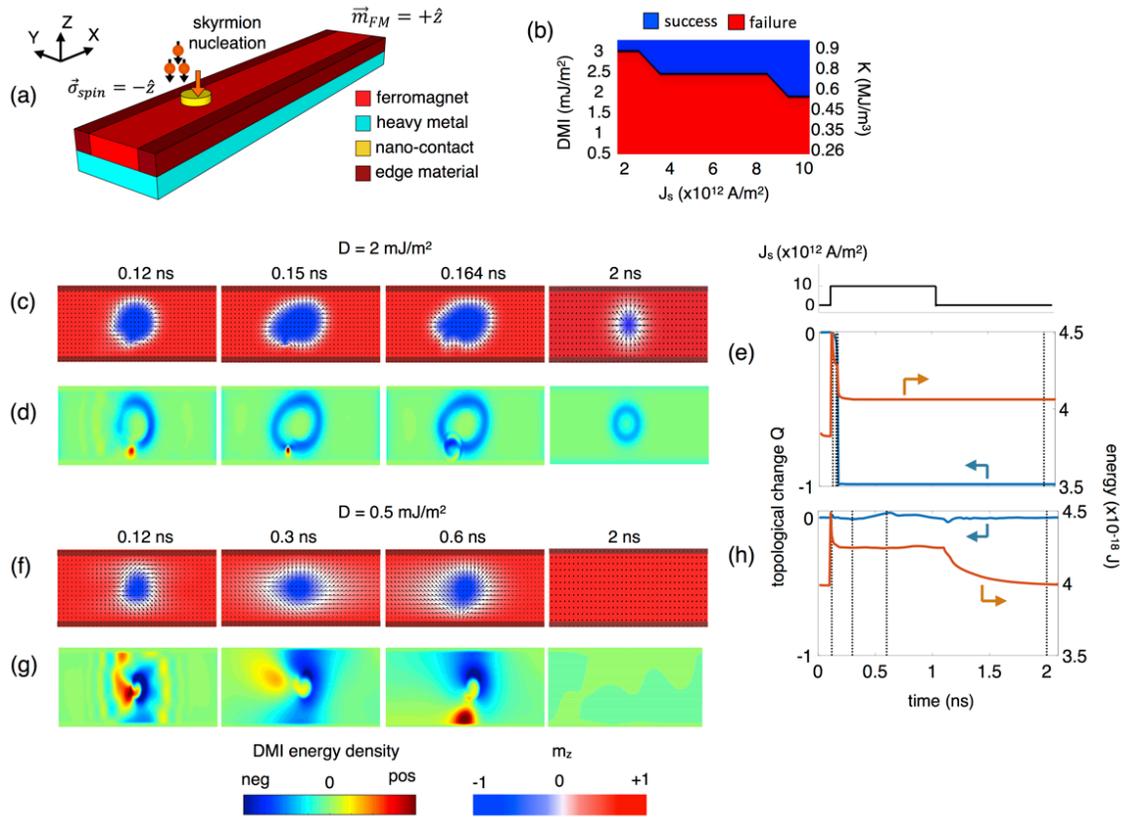

Figure 3. (a) Scheme of skyrmion nucleation in a nanowire with edge material exhibiting high PMA. (b) Probability of skyrmion nucleation in the presence of edge material for different values of (D, K) and injected spin current density $J_s$. (c, d) Temporal evolution of magnetization and corresponding DMI energy density after spin current injection for D = 2 mJ/m$^2$. (e) Change in the skyrmion number from 0 to -1 upon VBL annihilation and the change in the total energy. (f, g) Temporal evolution of magnetization and corresponding DMI energy density after spin current injection for D = 0.5 mJ/m$^2$. (h) Corresponding change in the skyrmion number and the total energy.

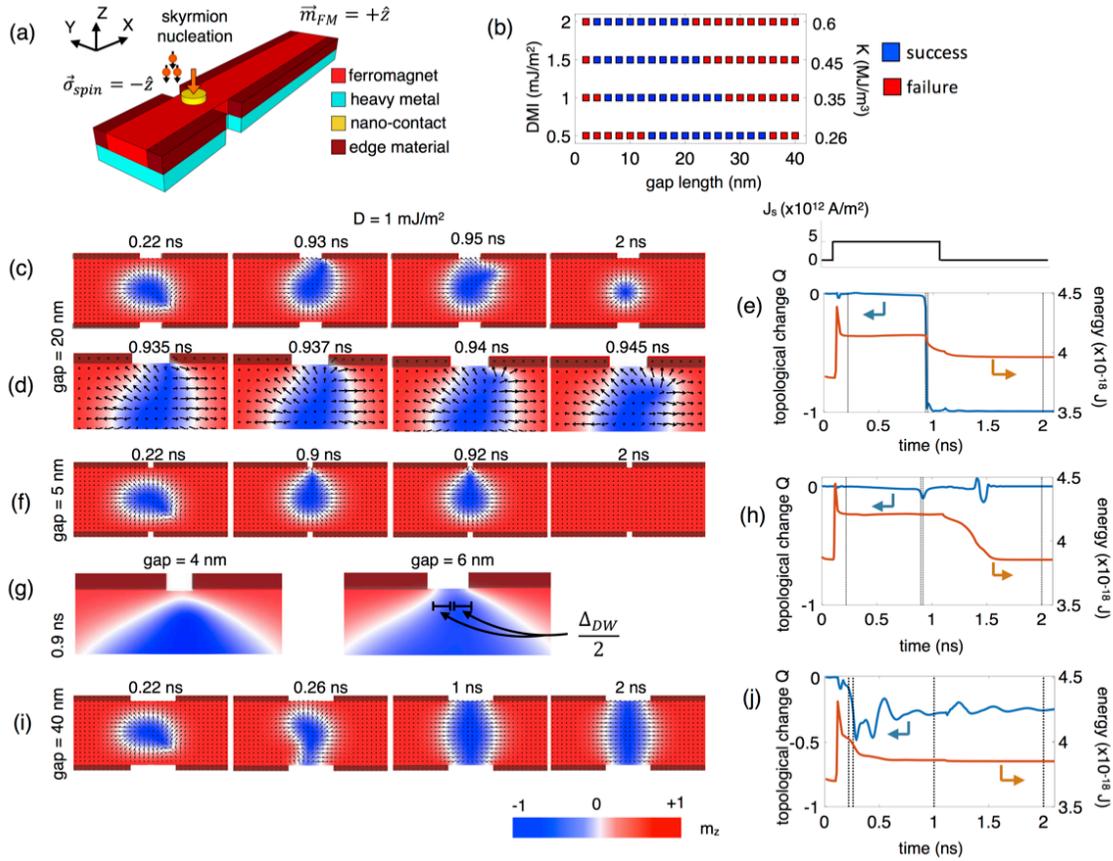

Figure 4. (a) Scheme of skyrmion nucleation using localized spin current injection into a nanowire with narrow gap in edge material. (b) Optimum gap length in the edge material for successful skyrmion nucleation for different values of DMI and K. (c) Temporal evolution of magnetization after spin current injection for D = 1 mJ/m². (d) Zoom-in of the gap region illustrating the skyrmion nucleation mechanism. (e) Change in the skyrmion number and the total energy for skyrmion nucleation process in the presence of a gap length of 20 nm. (f) Temporal evolution of magnetization in the presence of a narrow edge gap of length 5 nm showing an unsuccessful skyrmion nucleation. (g) Zoom-in of the gap region for two different gap lengths of 4 and 6 nm illustrating the requirement of a minimum gap length equal to the domain wall width $\Delta_{DW} = \sqrt{A/K_{eff}}$ for successful skyrmion nucleation. (h) Change in the skyrmion number and the total energy in the presence of a gap length of 5 nm. (i) Temporal evolution of magnetization in the presence of a wide edge gap of length 40 nm showing the formation of an edge-meron and finally a domain wall pair. (j) Change in the skyrmion number and total energy in the presence of a gap length of 40 nm.

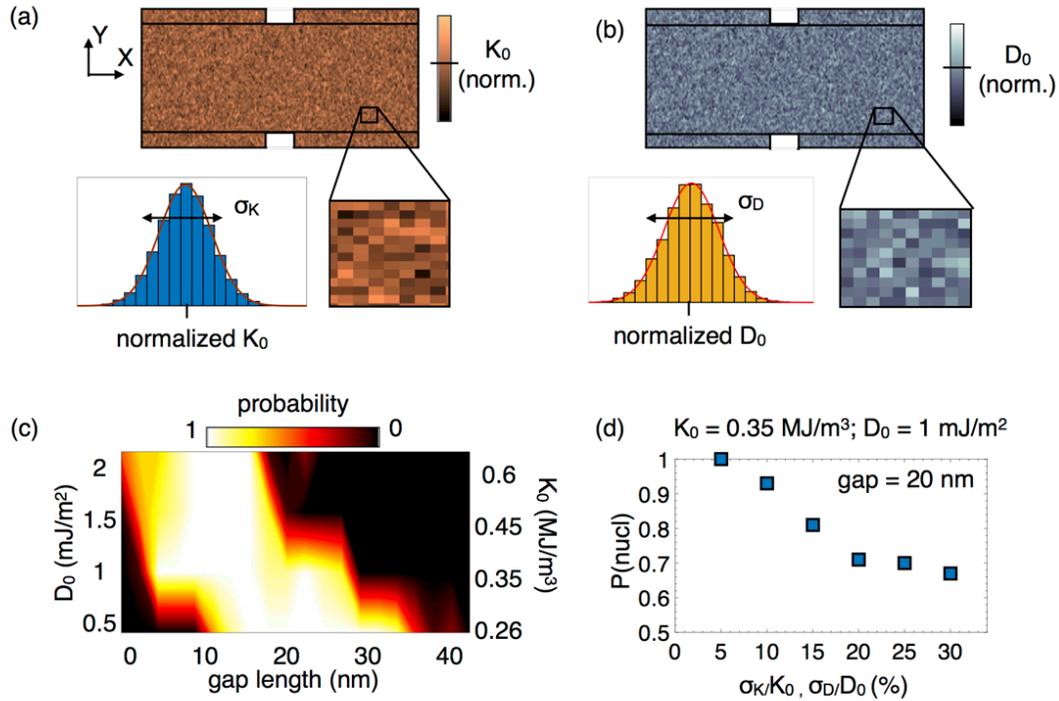

Figure 5. Scheme of a local site-by-site uncorrelated spatial variation of (a) PMA and (b) DMI used in the simulation. We use a normal distribution around a mean value of $K_0$ and $D_0$ with standard deviations ($\sigma_K$ and $\sigma_D$) of 10%. (c) Probability distribution plot of nucleating a skyrmion for different values of DMI (and PMA) and gap length, which agrees well with the variation-free plot in Fig. 3(j) illustrating the robustness of the proposed nucleation mechanism. (d) Probability of nucleation for simultaneous variation in PMA and DMI ($\sigma_K$ and $\sigma_D$) from 5-30% around the mean value of $K_0$ = 0.35 MJ/m$^3$ and $D_0$ = 1 mJ/m$^2$.

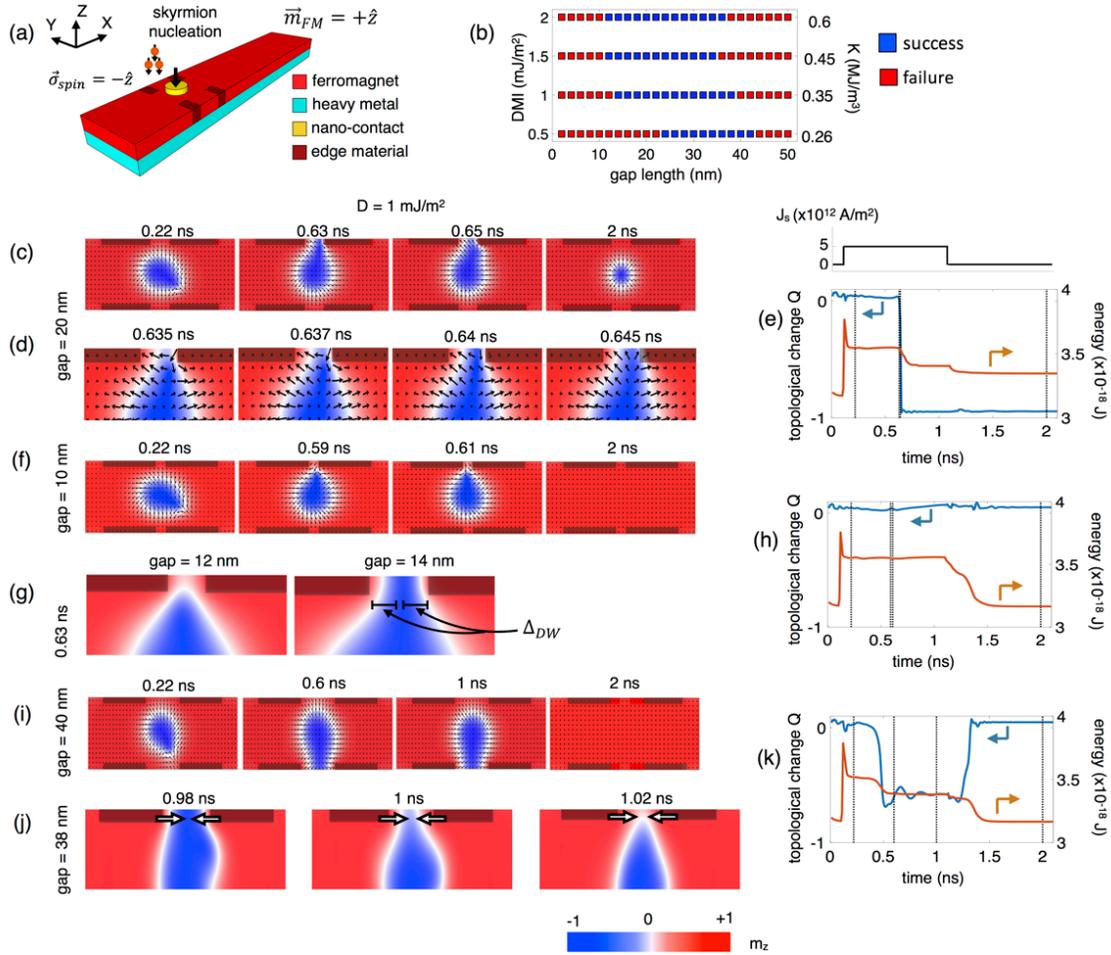

Figure 6. (a) Scheme of skyrmion nucleation in a nanowire with notches as pinning sites. (b) Optimum gap length between the notches for successful skyrmion nucleation for different values of DMI and K. (c) Temporal evolution of magnetization after spin-current injection for D = 1 mJ/m$^2$. (d) Zoom-in of the gap region between the notches illustrating the skyrmion nucleation mechanism. (e) Change in the skyrmion number and the total energy for skyrmion nucleation process in the presence of a gap length of 20 nm. (f) Temporal evolution of magnetization in the presence of a narrow gap of 10 nm between the notches showing an unsuccessful skyrmion nucleation. (g) Zoom-in of the gap region for two different gap lengths of 12 and 14 nm illustrating the requirement of a minimum gap length of $2\Delta_{DW}$ for successful skyrmion nucleation. (h) Change in the skyrmion number and the total energy in the presence of a gap length of 10 nm. (i) Temporal evolution of magnetization in the presence of a wide edge gap of length 40 nm showing the formation of an edge-meron. (j) Zoom-in of the gap region for a 38 nm gap length showing the additional repulsive force from the notches facilitating the generation of defect-free skyrmion bubble. (k) Change in the skyrmion number and the total energy in the presence of a gap length of 40 nm.